\documentclass[%
reprint,
superscriptaddress,
 amsmath, amssymb,
 aps, physrev,
 pra,
]{revtex4-2}

\usepackage{graphicx}% Include figure files
\usepackage{bm}% bold math
\usepackage[
 colorlinks=true,
 allcolors=blue,
 breaklinks=true,
]{hyperref}

\usepackage{dcolumn}
\usepackage[T1]{fontenc}
\usepackage{enumitem}

\begin{document}

\title{Measurement-Based Entanglement Distillation and Constant-Rate Quantum Repeaters over Arbitrary Distances}

\author{Yu Shi}
\email{shiyuphys@gmail.com}
\affiliation{Department of Electrical and Computer Engineering, University of Maryland, College Park, Maryland 20742, USA}
\affiliation{NSF-ERC Center for Quantum Networks, University of Arizona, Tucson, Arizona 85721, USA}
\affiliation{Wyant College of Optical Sciences, University of Arizona, Tucson, Arizona 85721, USA}

\author{Ashlesha Patil}
\affiliation{NSF-ERC Center for Quantum Networks, University of Arizona, Tucson, Arizona 85721, USA}
\affiliation{Wyant College of Optical Sciences, University of Arizona, Tucson, Arizona 85721, USA}

\author{Saikat Guha}
\email{saikat@umd.edu}
\affiliation{Department of Electrical and Computer Engineering, University of Maryland, College Park, Maryland 20742, USA}
\affiliation{NSF-ERC Center for Quantum Networks, University of Arizona, Tucson, Arizona 85721, USA}
\affiliation{Wyant College of Optical Sciences, University of Arizona, Tucson, Arizona 85721, USA}

\date{\today}

\begin{abstract}

Measurement-based quantum repeaters employ entanglement distillation and swapping across links using locally prepared resource states of minimal size and local Bell measurements. In this Letter, we introduce a systematic protocol for measurement-based entanglement distillation and its application to repeaters that can leverage any stabilizer code. Given a code, we explicitly define the corresponding resource state and derive an error-recovery operation based on all Bell measurement outcomes. Our approach offers deeper insights into the impact of resource state noise on repeater performance while also providing strategies for efficient preparation and fault-tolerant preservation of resource states. As an application, we propose a measurement-based repeater protocol based on quantum low-density parity-check (QLDPC) codes, enabling constant-yield Bell state distribution over arbitrary distances. Numerical simulations identify a fault-tolerant threshold on the total physical error per repeater segment---including errors on resource states, remotely generated Bell states, and Bell measurements---and confirm that increasing the QLDPC code size further suppresses the logical error while maintaining a fixed encoding rate. This work establishes a scalable backbone for future global-scale fault-tolerant quantum networks.

\end{abstract}

\maketitle

\emph{Introduction---}Entanglement plays a crucial role in quantum networks~\cite{Wehner2018, Awschalom2021, Azuma2023, Li2023}, quantum metrology~\cite{Giovannetti2011, Gottesman2012, Toth2014, Pezze2018}, and distributed quantum computing~\cite{Gottesman1999, Dur2003, Briegel2009, Li2012, Hu2023}. Establishing entanglement over long-range noisy channels requires quantum repeaters~\cite{Briegel1998, Duan2001, Hartmann2007, Jiang2009, Choe2024}, which correct noise and relay entanglement over intervals. Since entanglement distillation and quantum error correction correspond one-to-one~\cite{Shi2025, Wilde2007, Dur2007, Bennett1996}, repeaters based on both can be fundamentally unified using quantum error correction theory. Conventional circuit-based repeater architectures require a complex sequence of operations~\cite{Jiang2009}, including logical state preparation, entanglement distillation, teleportation-based gates, and logical Bell measurements. These steps demand deep, fault-tolerant circuits, additional ancilla qubits, and multiple rounds of error correction, leading to significant overhead and latency. In contrast, measurement-based repeaters~\cite{Zwerger2012, Zwerger2013, Zwerger2016, Zwerger2018, Yan2021, Yan2022, Raussendorf2001} rely solely on locally prepared resource states and single-shot Bell measurements, effectively resembling physical entanglement swapping. This approach eliminates the need for long sequences of coherent gates on memory qubits, thereby minimizing operational complexity, reducing latency, and mitigating error accumulation. A key advantage is that resource states can be prepared offline, enabling error detection and distillation prior to use. Moreover, measurement-based protocols offer greater flexibility, as resource states can be generated through more general procedures rather than being restricted to fixed unitary circuits. This adaptability makes the approach well suited to a broad range of physical platforms, including trapped ions~\cite{Moehring2007, Lanyon2013, Santra2019, Dhara2022, Kang2023, Krutyanskiy2023}, reconfigurable systems like neutral atoms~\cite{Bluvstein2024, Xu2024, Li2024, Pattison2024, Constantinides2024}, and architectures lacking direct qubit-qubit interactions, such as photons~\cite{Azuma2015, Huang2022, Huang2022a, Bartolucci2023, Patil2024} and solid-state quantum memories~\cite{Pichler2017, Buterakos2017, Zhan2020, Shi2021, Wan2021, Li2022, Liu2024, Stas2022, Knaut2024, Parker2024, Beukers2024}.

Zwerger et al.~\cite{Zwerger2012, Zwerger2013} proposed a measurement-based quantum repeater architecture by deriving resource states through the channel-state duality~\cite{Jamiołkowski1972} of unitary encoding circuits. However, this method lacks a systematic approach to resource state construction, making it difficult to extend to high-performance quantum codes and limiting its practical applicability. Later, they introduced a hashing-based protocol that achieves a constant yield over any distance~\cite{Zwerger2018}. While promising in theory, this proposal remains largely conceptual: it does not provide explicit resource states, and the required error decoding and state preparation are computationally and experimentally demanding. In contrast, modern quantum error-correcting codes, such as quantum low-density parity-check (QLDPC) codes~\cite{Breuckmann2021}, offer significant practical advantages. Their constant-weight stabilizers simplify resource state preparation and protection~\cite{Shi2025}, while their constant encoding rate and efficient decoding ensure scalability for long-distance communication. These features address key limitations of previous approaches and bring measurement-based quantum repeaters closer to practical realization.

In this Letter, we extend measurement-based entanglement distillation and repeater protocols to any stabilizer code. Using the parity-check matrix of a code, we explicitly define the resource states and present an error-recovery operation based on Bell measurement outcomes. These resource states correspond to physical-logical Bell states for distillation and logical-logical Bell states for repeaters. Their preparation simplifies to locally encoding logical Bell states, a process that can be implemented with constant circuit depth for certain codes, such as QLDPC codes. The resource states are preserved through syndrome error correction and entanglement distillation, offering greater efficiency than the graph state distillation approach~\cite{Dur2003a}. As an application, we propose a repeater protocol based on QLDPC codes, enabling constant-yield Bell state distribution over arbitrary distances, with efficient resource state preparation. Numerical simulations show that the protocol achieves fault tolerance when the combined error probability satisfies $p_r+p_B+p_m \le 4.65\%$, where $p_r$, $p_B$, and $p_m$ denote the error rates of local resource state qubits, remotely generated Bell state qubits, and Bell measurements, respectively. Our protocol ensures efficiency and scalability, laying the foundation for global-scale quantum networks.

\emph{Measurement-based entanglement distillation---}We consider an $[\![n,k,d]\!]$-quantum error-correcting code with a binary check matrix $\bm{H}=\left(\bm{H}_1\middle| \bm{H}_2\right)$~\cite{Nielsen2010}. The steps of the measurement-based entanglement distillation protocol based on this code are outlined below. We provide a schematic example in Figure~\ref{fig:distillation} for illustration and include the derivation of the protocol in Appendix~\ref{app:derivation}.

\begin{figure}[b]
    \centering
    \includegraphics[width=0.4\paperwidth]{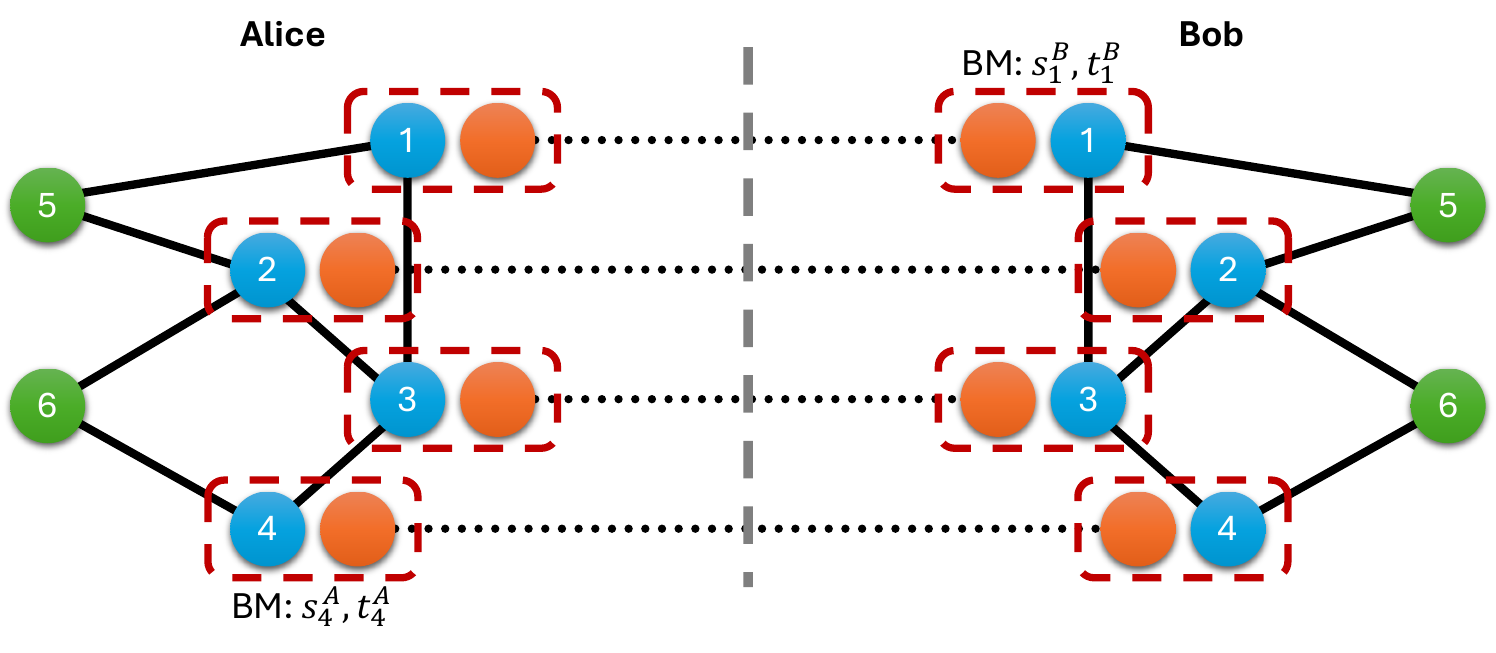}
    \caption{Schematic of measurement-based entanglement distillation using the $[\![4,2,2]\!]$ code with parity checks $X_1X_2X_3X_4$ and $Z_1Z_2Z_3Z_4$. The 6-qubit stabilizer states, drawn in graph states~\cite{Hein2004}, process 4 noisy Bell states into 2 distilled Bell states. Blue circles denote interfacing qubits, green circles denote output qubits, and orange circles denote input Bell states. Dashed red rectangles indicate Bell measurements (BMs) with parities $s_i^{A\left(B\right)}$ for $XX$ and $t_i^{A\left(B\right)}$ for $ZZ$.}
    \label{fig:distillation}
\end{figure}

Step 1: Alice and Bob prepare their resource states, i.e., an $\left(n+k\right)$-qubit stabilizer state with generators
\begin{equation} \label{eq:distillation_stabilizer}
    \left\{ g_i,\ \bar{X}_jX_{n+j},\ \bar{Z}_jZ_{n+j} \right\}\;.
\end{equation}
Here, $g_i$ represents the code stabilizer generator, where $i=1,2,\ldots,n-k$. The code has supports on the first $n$ qubits. $\bar{X}_j$ and $\bar{Z}_j$ represent the Pauli operators of the $j$-th logical qubit, $X_{n+j}$ and $Z_{n+j}$ represent the Pauli operators of the $\left(n+j\right)$-th physical qubit, where $j=1,2,\ldots,k$. $\bar{X}_jX_{n+j}$ and $\bar{Z}_jZ_{n+j}$ denote the tensor products $\bar{X}_j\otimes X_{n+j}$ and $\bar{Z}_j\otimes Z_{n+j}$, respectively. The resource state forms $k$ logical-physical Bell states, represented by
\begin{equation} \label{eq:distillation_state}
    \left|\mathcal{R}_D\right\rangle=\frac{1}{\sqrt{2^k}}\bigotimes_{j=1}^{k}{\left(\left|\bar{0}_j\right\rangle\left|0_{n+j}\right\rangle+\left|\bar{1}_j\right\rangle\left|1_{n+j}\right\rangle\right)}\;,
\end{equation}
where $\bar{0}_j$ and $\bar{1}_j$ represent the computational bases of the $j$-th logical qubit, and $0_{n+j}$ and $1_{n+j}$ represent the bases of the $(n+j)$-th physical qubit.

Step 2: Alice and Bob each receive $n$ input qubits, prepared as pairs in the standard Bell state
\begin{equation*}
    \left|\Phi^+\right\rangle=\frac{1}{\sqrt2}\left(\left|00\right\rangle+\left|11\right\rangle\right)\,,
\end{equation*}
but have undergone noise before their delivery. Alice performs Bell measurements on the qubits she received and the first $n$ qubits of her resource state. These measurements yield outcomes $\left(s_i^A,t_i^A\right)$, where $s_i^A$ is the parity of the $XX$-measurement and $t_i^A$ is the parity of the $ZZ$-measurement, for $i=1,2,\ldots,n$. Bob mirrors Alice's operations, obtaining outcomes $\left(s_i^B,t_i^B\right)$.

Step 3: Alice sends her measurement results to Bob, who calculates the overall parities, represented by vectors $\bm{s}$ and $\bm{t}$, with entries given by $s_i=s_i^A+s_i^B$ and $t_i=t_i^A+t_i^B$, respectively. Bob then computes the error syndrome by
\begin{equation} \label{eq:syndrome}
    \bm{S}=\bm{H}_1\cdot\bm{s} + \bm{H}_2\cdot\bm{t} + \bm{r}\;,
\end{equation}
where $\bm{r}$, with
\begin{equation*}
    r_i=\sum_{j=1}^{n}{H_1\left(i,j\right)H_2\left(i,j\right)}\;,
\end{equation*}
compensates for the effects of Pauli-$Y$ operators in the code stabilizers. Note that all addition in this protocol is performed modulo 2.

Step 4: Using the error syndrome $\bm{S}$, Bob employs a decoder to estimate the noise $\hat{\bm{e}}$, as in conventional quantum error correction, and identifies the effective unilateral bit and phase flips on his output qubits. The bit flips are calculated by
\begin{equation} \label{eq:bit_flip}
    \bm{\beta}=\left(\bm{Z},\hat{\bm{e}}\right)_{\rm sp}+\bm{Z}_1\cdot \bm{s} + \bm{Z}_2\cdot \bm{t} + \bm{r}^b\;,
\end{equation}
where $\bm{Z}=\left(\bm{Z}_1\middle| \bm{Z}_2\right)$ represents the binary matrix of logical $Z$ operators, and $\left(\bm{Z},\hat{\bm{e}}\right)_{sp}$ is the symplectic inner product~\cite{Calderbank1998}, representing the error-induced bit flips. The remaining terms represent the measurement-induced bit flips. The term $\bm{r}^b$, with 
\begin{equation*}
    r_i^b=\sum_{j=1}^{n}{Z_1\left(i,j\right)Z_2\left(i,j\right)}\;,
\end{equation*}
compensates for the effects of Pauli-$Y$ operators. The phase flips are computed similarly by
\begin{equation} \label{eq:phase_flip}
    \bm{\phi}=\left(\bm{X},\hat{\bm{e}}\right)_{\rm sp} + \bm{X}_1\cdot \bm{s} + \bm{X}_2\cdot \bm{t} + \bm{r}^p\;,
\end{equation}
where $\bm{X}=\left(\bm{X}_1\middle| \bm{X}_2\right)$ represents the binary matrix of logical X operators, and the elements of $\bm{r}^p$,
\begin{equation*}
    r_i^p=\sum_{j=1}^{n}{X_1\left(i,j\right)X_2\left(i,j\right)}\;. 
\end{equation*}

Step 5: Bob corrects the bit and phase flips by applying appropriate Pauli gates on his $\left(n+i\right)$-th qubit, transforming $\left(\beta_i,\phi_i\right)$ to $\left(0,0\right)$, for $i=1,2,\ldots,k$. The remaining qubits, indexed by $n+i$ for $i=1,2,\ldots,k$, on Alice's and Bob's sides form $k$ distilled Bell states, targeting the state $\left|\Phi^+\right\rangle$.

The above describes a one-way entanglement distillation protocol. A two-way protocol can also be implemented, in which Bob verifies only whether the syndrome $\bm{S}=\bm{0}$ to determine whether the distillation succeeds or fails, then communicates the result to Alice. This protocol can tolerate more errors (up to $d-1$) but is probabilistic and requires two-way classical communication.

\begin{figure*}[tb]
    \centering
    \includegraphics[width=0.8\paperwidth]{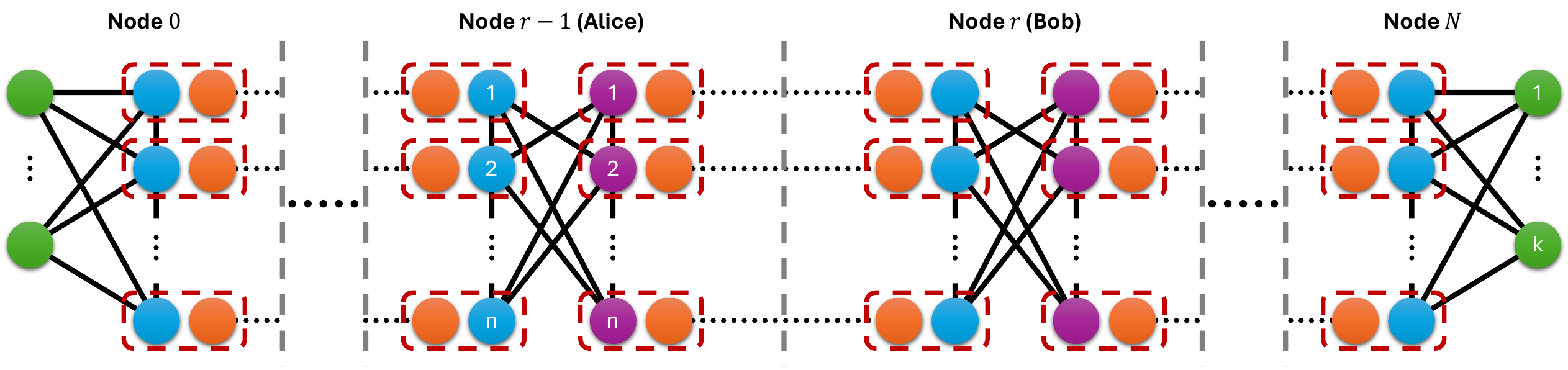}
    \caption{Schematic of high-fidelity Bell state distribution over long distances using measurement-based quantum repeaters. The total distance is divided into $N$ segments with $N+1$ nodes. Intermediate nodes perform Bell measurements (dashed red rectangles) on input Bell states (orange circles) and resource states (blue and purple circles) derived from an $[\![n,k,d]\!]$-quantum error-correcting code, relaying entanglement to establish Bell states (green circles) between the end nodes.}
    \label{fig:repeater}
\end{figure*}

\emph{Measurement-based quantum repeaters---}Distributing high-fidelity Bell states over long distances requires quantum repeaters, which can be implemented using a measurement-based approach, as illustrated in Figure~\ref{fig:repeater}. The total distance is divided into $N$ segments with $N+1$ nodes, where neighboring nodes share noisy Bell states (orange circles). Nodes $0$ and $N$ mark the end nodes, while intermediate nodes, labeled $1$ through $N-1$, function as quantum repeaters. Each intermediate node performs Bell measurements using a resource state (blue and purple circles) derived from an $\left[\!\left[n,k,d\right]\!\right]$-quantum error-correcting code. This enables the swapping of entanglement of $k$ logical qubits between neighboring nodes while applying error correction. Thus, the quantum repeaters relay the logical Bell states, establishing entanglement between nodes $0$ and $N$. At the end nodes, measurement-based entanglement distillation transforms the logical Bell states into physical Bell states (green circles).

We outline the repeater protocol below, considering two neighboring repeater nodes $r$ and $r+1$, also referred to as Alice and Bob, respectively, as shown in Figure~\ref{fig:repeater}.

Step 1: Alice and Bob prepare their resource states, a $2n$-qubit stabilizer state with generators
\begin{equation*}
    \left\{ g_i,\ g_i^\prime,\ \bar{X}_j\bar{X}_j^\prime,\ \bar{Z}_j\bar{Z}_j^\prime \right\}\;.
\end{equation*}
Here, $g_i$ and $g_i^\prime$, for $i=1,2,\ldots,n-k$, represent the code stabilizer generators, acting on the blue and green qubits, respectively. $\bar{X}_j\bar{X}_j^\prime$ and $\bar{Z}_j\bar{Z}_j^\prime$ denote the tensor products of Pauli operators for the corresponding $j$-th logical qubits, where $j=1,2,\ldots,k$. The resource state forms $k$ logical Bell states, represented by
\begin{equation} \label{eq:repeater_state}
    \left|\mathcal{R}_S\right\rangle=\frac{1}{\sqrt{2^k}}\bigotimes_{j=1}^{k}{\left(\left|{\bar{0}}_j\right\rangle\left|{\bar{0}}_j^\prime\right\rangle+\left|{\bar{1}}_j\right\rangle\left|{\bar{1}}_j^\prime\right\rangle\right)}\;.
\end{equation}

Steps 2--4: These steps mirror those in the distillation protocol, establishing logical Bell states between Alice’s blue qubits and Bob’s purple qubits. Bob then determines the bit flips $\bm{\beta}^{(r)}$ and phase flips $\bm{\phi}^{(r)}$ in the resulting state using Equations~\ref{eq:bit_flip} and~\ref{eq:phase_flip}, accounting for the effects of errors and measurements.

Step 5: Each node transmits its computed bit and phase flips to the end node $N$, which determines the overall flips as $\bm{\beta}=\sum_{r=1}^{N}\bm{\beta}^{(r)}$ and $\bm{\phi}=\sum_{r=1}^{N}\bm{\phi}^{(r)}$, with summation performed modulo $2$. Node $N$ then applies the appropriate Pauli gates to its green qubits to correct these accumulated errors, transforming $\left(\beta_i,\phi_i\right)$ to $\left(0,0\right)$ for $i=1,2,\ldots,k$. Consequently, $k$ high-fidelity physical Bell states are established between Nodes $0$ and $N$.

Quantum repeaters relay logical Bell states across nodes, analogous to storing logical quantum memories over time. Each segment effectively teleports logical Bell states while applying quantum error correction, akin to an error correction cycle for quantum memories. Given that neighboring nodes share noisy Bell states with fidelity $\mathcal{F}$, the infidelity $p=1-\mathcal{F}$ represents the physical qubit error. Quantum error correction reduces this, yielding a logical error $p_L$ dependent on $p$ and the chosen error-correcting code. The infidelity of the distributed Bell states between the end nodes, due to accumulated logical errors, is approximated by $\left(1-p_L\right)^N\approx 1-Np_L$ for small $p_L$.

\emph{Error analysis---}The measurement-based entanglement distillation and repeater protocols involve four quantum components and operations: resource states, input Bell states, Bell measurements, and recovery Pauli gates. To perform a full error analysis, we model errors in each as follows. Imperfect resource states are modeled as prepared stabilizer states, with each qubit subject to a depolarizing channel with error probability $p_r$. Noisy input Bell states are similarly modeled, with each qubit undergoing a depolarizing channel with error probability $p_B$. Bell measurement error is modeled as a depolarizing process with error probability $p_m$, where the $XX$ outcome, the $ZZ$ outcome, or both are flipped, each with probability $p_m/3$. Recovery Pauli gates are treated as classical processing that tracks the Pauli frame and are assumed to be noise-free.

These three error sources collectively affect the Bell measurements on the input and resource state qubits, resulting in a combined depolarizing error with probability $p_c=f\left(p_r,p_B,p_m\right)$, which may flip the measurement outcomes and can be detected by the computed syndrome. For convenience, we define a function for composing two independent depolarizing errors as $f\left(p_1,p_2\right)=p_1+p_2-\frac{4}{3}p_1p_2$, which itself is a depolarizing error. This extends to multiple sources via $f\left(p_r,p_B,p_m\right)=f\left[f\left(p_r,p_B\right),p_m\right]$. The error probability $p_c$ represents a single-partite error between Alice and Bob, while the total bipartite error is given by $p_t=f\left(p_c,p_c\right)$. A single-shot error correction can correct all these errors because the Bell states and Bell measurements are symmetric, and the effects of these errors are degenerate. Therefore, the total error can be equivalently modeled as a depolarizing error, resulting in a shared Bell state with fidelity $\mathcal{F}=1-p_t$~\cite{Zwerger2013, Zwerger2018, Shi2025}. The parameter $p_t$ also characterizes the error per repeater segment. Error correction at each repeater node reduces it to a logical error $p_L$, which depends on $p_t$ and the chosen error-correcting code. Logical errors accumulate over $N$ repeater segments. The fidelity of the final distributed Bell states is approximated as $1-Np_L-2p_r$, where the $2p_r$ term accounts for uncoded output qubit errors at the end nodes. These errors are independent of transmission distance and can be mitigated through local resource state distillation.

Since the resource states form logical-physical and logical-logical Bell states, their preparation simplifies to encoding logical Bell states. A promising approach~\cite{Shi2025} is to locally prepare $n$ copies of physical Bell states first and then use bilateral stabilizer measurements to encode them onto logical Bell states, forming the repeater resource state. To obtain the distillation resource state, one can decode one part of the logical Bell state using single-qubit measurements. For certain codes, such as QLDPC codes, this preparation method features a constant circuit depth. Preserving the distillation resource state involves two components. The logical subsystems are protected via syndrome error correction. The physical systems can be stabilized using $\bar{X}X$ and $\bar{Z}Z$ measurements. However, as their weight increases with code distance, these measurements become inefficient. A more practical approach is to locally distill logical-physical Bell states~\cite{Shi2025}, avoiding large-weight stabilizer measurements. Preserving the repeater resource state is more straightforward, as the logical subsystems can be independently corrected by syndrome error correction.

\emph{Quantum repeaters with constant yield rate---}We propose a measurement-based quantum repeater protocol using QLDPC codes for fault-tolerant Bell state distribution at a constant yield rate over any distance. These codes have a constant number of qubits per stabilizer generator and a constant number of generators per qubit, enabling efficient resource state preparation~\cite{Shi2025}. They also maintain a constant encoding rate and good code distance, exponentially suppressing logical errors~\cite{Gottesman2014, Rengaswamy2024}. For $\left[\!\left[n,k=rn,d=O\left(\sqrt n\right)\right]\!\right]$-hypergraph product (HGP) codes with constant rate $r$~\cite{Tillich2014}, the logical error is given by~\cite{Xu2024}
\begin{equation} \label{eq:fitting}
    p_L=c\left(\frac{p}{p_0}\right)^{\alpha n^\delta}\;,
\end{equation}
where $p$ is the physical error, and $p_0$ is the error threshold. The constant $c$ serves as an overall prefactor, while the exponent $\alpha n^\delta$ characterizes the scaling of the code distance with the number of physical qubits in a given family of quantum codes~\cite{Gottesman2014}. For the HGP codes, the distance asymptotically scales as $d\propto\sqrt n$, corresponding to $\delta=\frac{1}{2}$. To distribute Bell states over distance $L$, we divide it into segments of length $L_0$, keeping neighboring-node Bell state infidelity below threshold. Each node then uses $2n$-qubit resource states to fault-tolerantly relay the entanglement. Consequently, the end-to-end Bell state infidelity is
\begin{equation*}
    \frac{cL}{L_0}\left(\frac{p}{p_0}\right)^{\alpha n^\delta}\;.
\end{equation*}
Any target fidelity can be achieved by increasing the code size $n$ while keeping the encoding rate $r$ fixed, with $n$ scaling polylogarithmically as $n=O(\log^{1/\delta}{L})$.

We simulate the repeater protocol using a family of HGP codes to compute the logical error $p_L$ as a function of the total physical error $p_t$ per repeater segment, as defined in the \textit{Error analysis} section. The logical error accumulates across repeater nodes and determines how far Bell states can be reliably distributed. The corresponding resource states can be efficiently prepared using a protocol developed in our previous work~\cite{Shi2025} on reconfigurable neutral-atom platforms. Following the procedure outlined in previous literature~\cite{Roffe2020, Grospellier2021, Xu2024}, we construct HGP codes by taking the hypergraph product of classical $(3,4)$-regular LDPC codes~\cite{Tillich2014}, which are randomly generated using software~\cite{Prabhakar2020}. From the generated instances, we select the code with the maximal distance. By increasing the code size, we obtain a family of HGP codes with a constant encoding rate of $0.04$. Then, we simulate the protocol using Stim~\cite{Gidney2021} with belief propagation-ordered statistics decoding (BP-OSD)~\cite{Roffe2022, Roffe2020, Panteleev2021}.

\begin{figure}[tb]
    \centering
    \includegraphics[width=0.4\paperwidth]{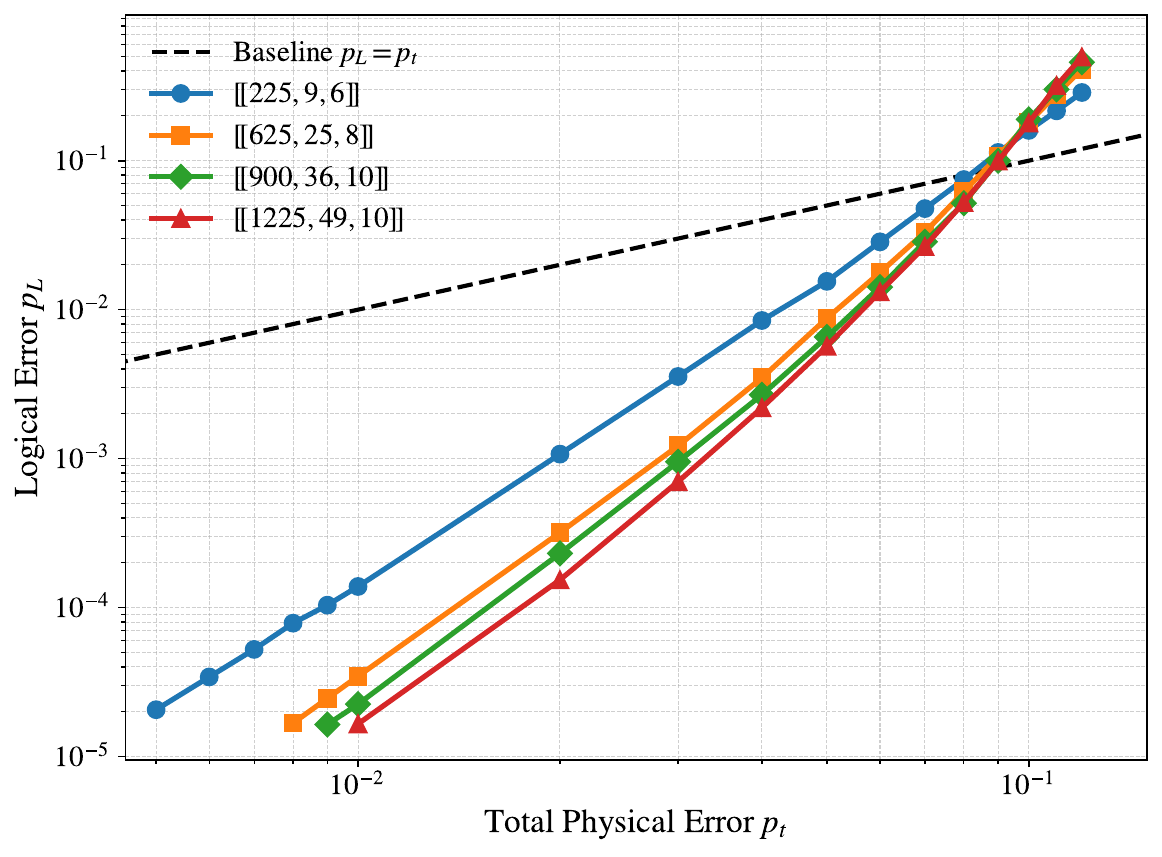}
    \caption{Logical error $p_L$ versus total physical error $p_t$ per repeater segment for the measurement-based repeater protocol using a family of hypergraph product (HGP) codes with a fixed encoding rate of $0.04$.}
    \label{fig:infidelity}
\end{figure}

Figure~\ref{fig:infidelity} shows $p_L$ versus $p_t$ for various code sizes. Because of finite-size effects in randomly generated codes, the $1225$-code and $900$-code share the same code distance. While it may seem unintuitive that the $1225$-code outperforms the $900$-code despite its larger size and identical distance, this is because error-correcting performance depends not only on the minimum weight but also on its overall distribution across codewords. We fit the data using Equation~\ref{eq:fitting} and extract a threshold value of $p_t=9.3\%$, implying fault-tolerant operation when the sum of individual errors satisfies $p_r+p_B+p_m \le 4.65\%$, where $p_r$, $p_B$, and $p_m$ are the error probabilities of resource state qubits, input Bell state qubits, and Bell measurements, respectively. The fidelity of input Bell states is given by $\mathcal{F}=1-2p_B+\frac{4}{3}p_B^2$. Additionally, increasing the code size further suppresses the logical error while maintaining a constant encoding rate. These results demonstrate that HGP codes support fault-tolerant, constant-rate Bell state distribution over arbitrary distances.

\emph{Conclusion and open questions---}In this Letter, we generalized measurement-based entanglement distillation and quantum repeaters to any stabilizer code. Given a code, we explicitly defined the corresponding resource states and presented an error-recovery procedure based on Bell measurement outcomes. We also analyzed the distillation capability of the protocol using noisy resource states and methods for efficiently preparing and preserving them. As an application, we proposed a quantum repeater protocol based on QLDPC codes that enables constant-rate Bell state distribution over arbitrary distances. Our simulations identified a fault-tolerant threshold on the total physical error for the proposed protocol. This work lays the foundation for scalable global quantum networks. We note that a related study on entanglement distillation using QLDPC codes~\cite{Ataides2025} was posted several days before this work; our study was conducted independently.

Several open questions remain for future research. One is implementing measurement-based entanglement distillation on real quantum platforms, where a key challenge is constructing resource states under constrained qubit connectivity. Another is designing quantum repeaters that specifically address photon loss. While loss can be heralded, enabling more efficient error correction, Bell measurements using linear optics remain limited by a $50\%$ success probability. Finally, achieving the hashing bound in quantum repeaters is an open challenge, potentially addressable through the quantum version of modern codes such as LDPC, turbo, or polar codes.

\begin{acknowledgments}
This research was supported by the National Science Foundation (NSF), under the following awards: ``Quantum Networks to Connect Quantum Technology (QuanNeCQT)'' awarded under grant 2134891, ``CIF-QODED: Quantum codes Optimized for the Dynamics between Encoded Computation and Decoding using Classical Coding Techniques'' awarded under grant 2106189, and the ``NSF Engineering Research Center for Quantum Networks (CQN)'' awarded under grant 1941583.
\end{acknowledgments}

\emph{Data availability---}The data that support the findings of this Letter are openly available~\cite{Dataset2025}.

\appendix

\section{Derivation} \label{app:derivation}
\begin{figure}[tb]
    \centering
    \includegraphics[width=0.4\paperwidth]{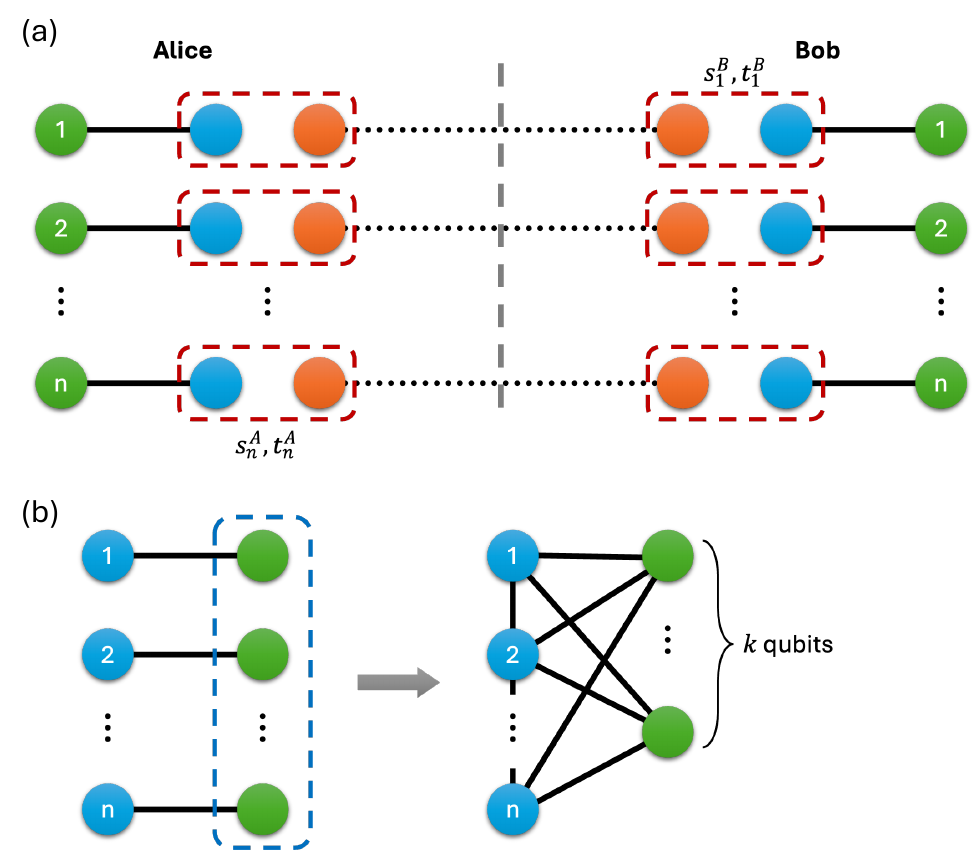}
    \caption{(a) Schematic of $n$ entanglement swapping processes, transferring entanglement from orange qubits shared between Alice and Bob to green qubits via Bell measurements. (b) Deriving the resource state by applying stabilizer and single-qubit measurements on the green qubits of local Bell states. The resulting states form logical-physical Bell states, with the blue qubits encoding $k$ logical qubits.}
    \label{fig:derivation}
\end{figure}

To derive the measurement-based entanglement distillation and quantum repeater protocols, we consider a setup consisting of $n$ rows of qubits, as shown in Figure~\ref{fig:derivation}(a). Each row represents an entanglement swapping process, where the connected green and blue circles represent local Bell states, and the orange circles represent a shared Bell state between Alice and Bob distributed over a noisy channel, all initialized in the standard Bell state
\begin{equation*}
    \left|\Phi^+\right\rangle=\frac{1}{\sqrt2}\left(\left|00\right\rangle+\left|11\right\rangle\right)\;.
\end{equation*}
Alice and Bob perform Bell measurements on the blue and orange qubits, yielding outcomes $s_i^{A\left(B\right)}$ for the $XX$-parity and $t_i^{A\left(B\right)}$ for the $ZZ$-parity, where $i$ labels the row. Through these operations, the entanglement is swapped from the orange qubits to the green qubits, subject to local Pauli corrections. Using the stabilizer formalism~\cite{Gottesman1997}, the standard Bell state can be represented by stabilizer generators $\left\{X\otimes X,Z\otimes Z\right\}$. After the measurements, the entangled green qubits are described by the stabilizers
\begin{equation} \label{eq:post_swap_stabilizer}
    \left\{\left(-1\right)^{s_i}X_i\otimes X_i,\ \left(-1\right)^{t_i}Z_i\otimes Z_i\right\}\;,
\end{equation}
where $s_i=s_i^A+s_i^B$, $t_i=t_i^A+t_i^B$, and $i=1,2,\ldots,n$. The $\otimes$ symbol indicates the bilateral tensor product, with $O_A\otimes O_B$ denoting operators $O_A$ and $O_B$ acting on Alice's and Bob’s sides, respectively.

We distill the $n$ entangled states formed by the green qubits into $k$ output states in the standard Bell states using an $[\![n,k,d]\!]$ quantum error-correcting code. The distillation method follows the stabilizer-based entanglement distillation protocol described in detail in~\cite{Shi2025}. This involves: (1) encoding the $n$ entangled states into $k$ logical Bell states through bilateral stabilizer measurements, and (2) decoding the logical states back into physical states using single-qubit measurements.

Given the $n$ entangled states described in Equation~\ref{eq:post_swap_stabilizer}, Alice and Bob perform $n-k$ stabilizer measurements on their qubits, resulting in outcomes $a_i$ and $b_i$, respectively. These bilateral stabilizer measurements project the $n$ entangled states into $k$ logical entangled states, characterized by stabilizers
\begin{equation} \label{eq:repeater_stabilizer}
\left\{ \begin{aligned}
&\left(-1\right)^{a_i} g_i \otimes I, \\
&\left(-1\right)^{b_i} I \otimes g_i, \\
&\left(-1\right)^{\phi_j} \bar{X}_j \otimes \bar{X}_j, \\
&\left(-1\right)^{\beta_j} \bar{Z}_j \otimes \bar{Z}_j
\end{aligned} \right\}\;,
\end{equation}
for $i=1,2,\ldots,n-k$ and $j=1,2,\ldots,k$. The measurement outcomes $a_i$ and $b_i$ are random because the code stabilizers $g_i\otimes I$ and $I\otimes g_i$ anti-commute with $X_j\otimes X_j$ or $Z_j\otimes Z_j$ for any $j$-th qubit within the support of the stabilizer. However, their sum, representing the parity of $g_i\otimes g_i$, is fixed and determined by a linear combination of $X_j\otimes X_j$ and $Z_j\otimes Z_j$ over the supports of $g_i\otimes g_i$, expressed as
\begin{equation} \label{eq:parity_identity}
    a_i+b_i=\bm{H}_1\left(i\right)\cdot\bm{s} + \bm{H}_2\left(i\right)\cdot\bm{t} + \bm{H}_1\left(i\right)\cdot \bm{H}_2\left(i\right)^T\;,
\end{equation}
where $\left(\bm{H}_1\left(i\right)\middle| \bm{H}_2\left(i\right)\right)$ is the binary vector representation of $g_i$ (the $i$-th row of the parity check matrix $\bm{H}$), and $\bm{H}_1\left(i\right)\cdot \bm{H}_2\left(i\right)^T$ accounts for the number of Pauli-$Y$ components in $g_i$. These components contribute additional $-1$ due to $Y\otimes Y=-XZ\otimes XZ$. Since $a_i$ and $b_i$ are random, Alice and Bob post-select on both outcomes being $0$, enforcing that the right-hand side (RHS) of Equation~\ref{eq:parity_identity} equals $0$. Errors in the transmitted orange qubits can flip Bell measurements $\bm{s}$ and $\bm{t}$, altering the value of the RHS in Equation~\ref{eq:parity_identity}. This change reveals the error syndrome, as described in Equation~\ref{eq:syndrome}. Similarly, Equations~\ref{eq:bit_flip} and~\ref{eq:phase_flip} are derived to compute the parities $\phi_j$ of ${\bar{X}}_j\otimes{\bar{X}}_j$ and $\beta_j$ of ${\bar{Z}}_j\otimes{\bar{Z}}_j$ caused by Bell measurements and errors. So far, the derivation has led to a quantum repeater that fault-tolerantly teleports logical Bell states. For entanglement distillation, it is necessary to decode the logical states into physical states.

Given the logical Bell states in Equation~\ref{eq:repeater_stabilizer}, Alice and Bob can decode the logical states into physical states by performing $n-k$ single-qubit measurements~\cite{Shi2025}. For an $[\![n,k,d]\!]$ stabilizer code, the measurement bases are chosen to map the logical operators $\bar{X}_j$ and $\bar{Z}_j$ to the physical operators $\left(-1\right)^{u_j}X_j$ and $\left(-1\right)^{w_j}Z_j$, respectively, for $j=1,2,\ldots,k$. The phases $u_j$ and $w_j$ are linear functions of the measurement outcomes and can be set to $0$ by post-selection. The decoded states are stabilized by
\begin{equation*}
    \left\{\left(-1\right)^{\phi_j}X_j\otimes X_j,\ \left(-1\right)^{\beta_j}Z_j\otimes Z_j\right\}\;.
\end{equation*}
Bob then applies Pauli gates to correct any residual phases, yielding $k$ standard Bell states.

The resource state remains to be derived. As shown in Figure~\ref{fig:derivation}, entanglement swapping and distillation are performed on different qubits, allowing their order to be interchanged. The resource state is prepared by applying stabilizer and single-qubit measurements on the green qubits, which initially forms standard Bell states with the blue qubits. Post-selecting measurement outcomes of all $0$s yields the resource state, with its stabilizers defined in Equation~\ref{eq:distillation_stabilizer}. Importantly, this post-selection is purely conceptual; in practice, the resource state can be prepared deterministically according to its stabilizers.

\newpage
\bibliography{main.bib}

\end{document}